# Efficient photon capture on germanium surfaces using industrially feasible nanostructure formation

Kexun Chen[1], Joonas Isometsä[1], Toni P. Pasanen[1], Ville Vähänissi[1] and Hele Savin[1]

[1] Department of Electronics and Nanoengineering, Aalto University, Tietotie 3, 02150 Espoo, Finland

E-mail: kexun.1.chen@aalto.fi

**Abstract**

Nanostructured surfaces are known to provide excellent optical properties for various photonics devices. Fabrication of such nanoscale structures to germanium (Ge) surfaces by metal assisted chemical etching (MACE) is, however, challenging as Ge surface is highly reactive resulting often in micron-level rather than nanoscale structures. Here we show that by properly controlling the process, it is possible to confine the chemical reaction only to the vicinity of the metal nanoparticles and obtain nanostructures also in Ge. Furthermore, it is shown that controlling the density of the nanoparticles, concentration of oxidizing and dissolving agents as well as the etching time plays a crucial role in successful nanostructure formation. We also discuss the impact of high mobility of charge carriers on the chemical reactions taking place on Ge surfaces. As a result we propose a simple one-step MACE process that results in nanoscale structures with less than 10% surface reflectance in the wavelength region between 400 nm and 1600 nm. The method consumes only a small amount of Ge and is thus industrially viable and also applicable to thin Ge layers.

Keywords: nanostructures, nanoparticles, germanium, metal assisted chemical etching, photonics

## 1. Introduction

Germanium (Ge), which belongs to group IV like silicon (Si), has gained considerable attention as the next-generation material for photonics, wireless communication and optical semiconductor devices due to its high carrier mobility and superior absorption properties [1-4]. More specifically, Ge enables photon absorption in a much wider wavelength range than Si since Ge has a narrower bandgap (~0.67 eV) [5] than that of Si (~1.12 eV) [6]. Furthermore, due to the presence of direct bandgap transitions around 0.8 eV in Ge, only a thin layer is needed for efficient photon absorption. Despite of all these benefits, Ge devices are still today in infancy as compared to Si. One good example is how the surface reflectance and light trapping are controlled in state-of-the-art devices. Ge devices rely mostly on the conventional antireflection coatings with limited performance [7-9], while there are much more advanced technologies developed for Si that are based on nanotexturing the surfaces. Such texturing allows superior light absorption in a much wider wavelength range and also at large incidence angles [10]. There is clearly a need to have similar technologies developed for Ge as well since efficient photon capture is a pre-requisite for a high-performance device in almost all photonics and optical applications.

There are several different approaches developed for nanotexturing a Si surface; these are based on either femtosecond lasers [11], reactive ion etching [12] or metal nanoparticle assisted etching [13-16]. Out of these, the latter one, also known as metal assisted/catalyzed chemical etching (MACE, MCCE), has the highest industrial potential as it can be fully performed in normal wet benches without a need for any additional equipment making the process straightforward and low-cost. Therefore, it is no surprise that this technology has already found its way to production even in highly price-competitive Si photovoltaics [17-19].

In the case of Si, MACE process has enabled a drastic reduction of surface reflectance, even below 5%, within a wavelength range of 300-1100 nm [20]. In Ge, on the other hand, the MACE process is far less developed and there are only few works available [21-24]. The obtained reflectances are far from what has been achieved for Si and the best reported reflectance value is in the order of ~20% [22]. It seems that further reduction of the surface reflectance using MACE process has remained a challenge for Ge wafers. The root cause for the poor reflectance can be understood if we consider the chemistry behind the ideal MACE process. The metal nanoparticles (e.g. Pt [25,26], Au [27], Ag [28,29], Cu [30]) that are deposited on top of the wafer surface catalyze oxidation reaction so that the oxidation of the semiconductor takes place only under the metal nanoparticles. The preferential oxidation is then followed by subsequent dissolution with the etchant present in the same solution. This continuous process results in semiconductor consumption only under the nanoparticles resulting in the nanospike formation. Such process works indeed well in Si where the common MACE oxidant, $H_2O_2$, is reduced only within the vicinity of the metal nanoparticles. However, in the case of Ge, the reactivity with the common oxidizing agents is much higher, which results in fast oxidation of Ge surfaces even far away from the metal catalyst. Therefore, controlling the dimensions of the surface structures by MACE becomes much more difficult in Ge. Indeed, the reported MACE structures have appeared as shallow pits with micron scale resulting in only modest optical performance [22].

Recent studies have tried to tackle the fast oxidation rate of Ge surfaces to obtain more nanoscale structures with MACE. Both Ito et al. [21] as well as Lee et al. [22] suggested that by reducing the $H_2O_2$ concentration in the MACE solution, oxidation of Ge could be suppressed so that $H_2O_2$ would react only in the vicinity of the metal nanoparticles. However, in their studies Ag nanoparticles agglomerated to large clusters making nanostructure formation impossible. Lee et al. [22] had also troubles using HF as an etchant, thereby, they ended up proposing just diluted $H_2O_2$ for the structure formation since pure water is also known to dissolve oxidized Ge similar to HF but with a lower rate. Nevertheless, the solution without any HF etchant may not be industrially relevant as the process time extends easily to hours. Kawase et al. [23,24] had a different approach: they dissolved oxygen into water instead of using the common $H_2O_2$ oxidizing agent. Rezvani et al. [31], on the other hand, combined electrochemical etching to MACE by applying external bias voltage to Ge substrate in order to enhance current transport in the vicinity of the metal nanoparticles. While they both claimed nanostructure formation after such process modifications, neither optical nor electrical characterization was reported. Moreover, both of the proposed processes make the original MACE process much more complicated counterbalancing the original benefits of MACE (i.e. simplicity and low-cost).

In this paper, our goal is to address the above challenge of Ge nanostructure formation by developing a simple MACE process where the Ge oxidation is confined to the vicinity of the metal nanoparticles and the enlargement of the patterns is limited by fast enough etching. The process is optimized by controlling the amount of $H_2O_2$ concentration, Ag nanoparticle concentration and the etching time. Consequently, we propose a simple one-step MACE process and study the resulting Ge nanostructure morphology with Scanning Electron Microscope (SEM). Furthermore, we characterize the optical properties of the nanostructures with wavelengths ranging from 400 nm to 1600 nm using UV-Vis-NIR spectroscopy. Finally, the industrial applicability and other benefits of the developed process are discussed.

## 2. Experimental details

The substrates used in this study were antimony-doped (n-type) 4" single-side polished Ge wafers with (100) orientation, a thickness of 175 μm and a resistivity of 0.01 - 0.4 Ωcm. Prior to the MACE process, the wafers were dipped in 0.5% HF solution for 20 s to remove any possible native oxide. The HF dip was followed by a subsequent few second DI-water (DIW) rinse. One wafer was kept as a reference experiencing no treatments and is later called as an as-received wafer.

After preliminary tests we replaced the conventional multi-step MACE process with a single solution that combines the metal nanoparticle deposition and the nanostructure formation. The solution consisted of the following mixture: HF (50 wt%) : $H_2O$ : $H_2O_2$ (30 wt%) = 49 : 198 : 10 with 0.0004 mol/l of $AgNO_3$. We noticed that by using $AgNO_3$ concentration that is much lower than conventionally used in MACE, it is possible to prevent Ag nanoparticle agglomeration to large clusters, which finally enables narrower structure formation. In order to better understand the role of the oxidant, the concentration of $H_2O_2$ was then varied so that the so-called ρ values (molar ratio of HF/(HF+ $H_2O_2$)) with different $H_2O_2$ concentration were 93%, 95%, 97% and 99%. The etching time was also varied with specific intervals ranging from 3 min to 12 min. Finally, the samples were rinsed in DIW. The resulting structures were characterized with SEM and with integrating sphere-based UV-Vis-NIR spectroscopy.

## 3. Results and discussion

Since $H_2O_2$ plays a crucial role in the oxidation of Ge, we study first how the concentration of $H_2O_2$ affects the surface morphology after the one-step MACE process. Figure 1 shows cross-sectional SEM images of Ge surfaces after immersing the wafers for 3 min in MACE solution with various $H_2O_2$ concentrations. It is seen that the process works surprisingly well and the surface nanostructures are indeed formed with this 1-step MACE solution. Furthermore, the figure shows that the $H_2O_2$ concentration clearly affects the size of the formed surface structures. The highest ρ value (Figure 1(a)) produces quasi-nano Ge structure with a diameter of approximately 200 nm while the increase in $H_2O_2$ concentration gradually increases the diameter all the way to 1300 nm (Figure 1(b) to (d)). This tendency is in agreement with the earlier results [21,22], where increasing $H_2O_2$ concentration resulted in the enlargement of the structure diameters and the formation of inverted cone-shaped pores.

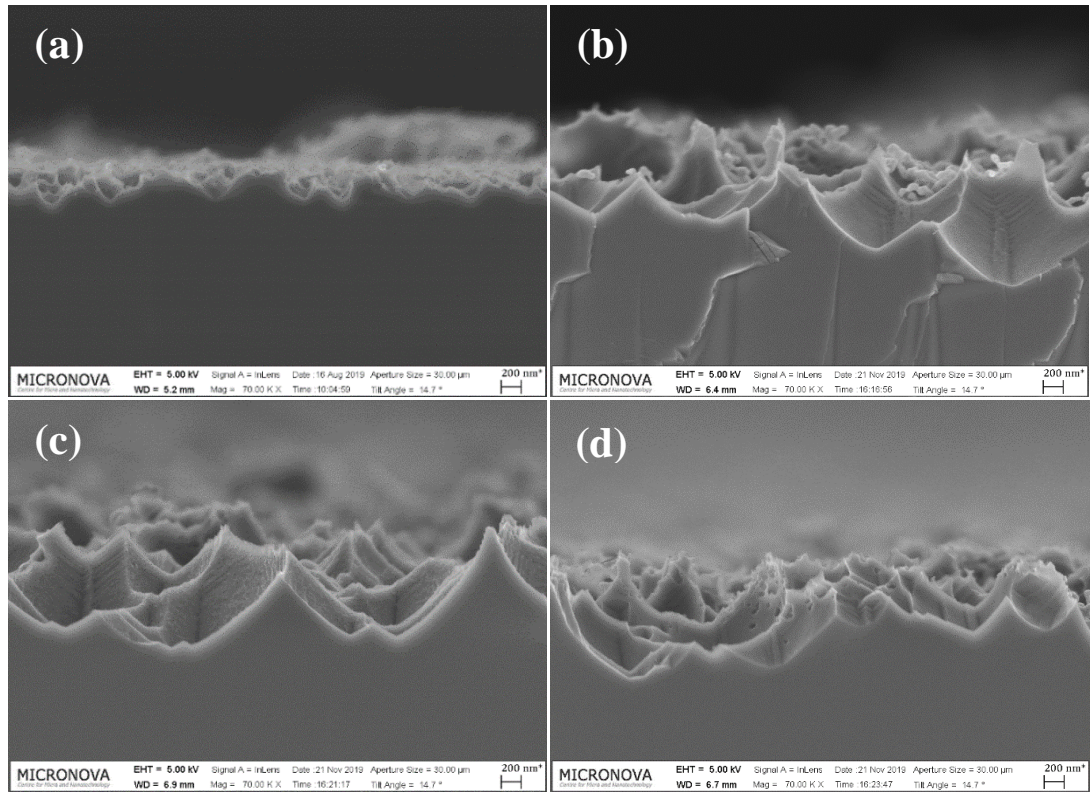

Figure 1. Cross-sectional SEM images of nanostructured Ge wafers etched for 3 min in 1-step MACE solution with different ρ values: (a) 99%, (b) 97%, (c) 95%, (d) 93%.

The SEM images indicate that in the case of high oxidant concentration, the dimensions of the etch pits are no longer determined by the small metal nanoparticles. This phenomenon was associated by Lee et al. [22] and Ito et al. [21] to the earlier observation that the Ge surface reacts aggressively with $H_2O_2$/HF even without the presence of Ag nanoparticles and thus the corresponding oxidation/etching takes place also on bare Ge surfaces. However, this behavior could also be related to the high charge-carrier mobility present in Ge substrates. Positively-charged carriers (holes) are generated in Ge due to $H_2O_2$ decomposition in the vicinity of the Ge/Ag interface and the amount of generated holes is proportional to the $H_2O_2$ concentration. In the case of high $H_2O_2$ and low HF concentration, the excess holes may not be consumed by HF with fast enough rate emphasizing the role of both the oxidant concentration as well as the etchant concentration. Subsequently, if the excess holes are not consumed near the nanoparticles, they manage to diffuse far away from the nanoparticles accelerating the reaction also outside the nanoparticles. The high mobility of holes makes this phenomenon much more severe in Ge than in Si. In reality, both the general reactivity of bare Ge surfaces and the diffusion of excess carriers and their interplay are possible explanations for the observed non-ideal MACE process with the given chemical compositions.

To investigate the optical performance of the same samples experiencing various ρ values, the reflectance spectra at a wavelength range of 400-1600 nm is presented in Figure 2. First, we can see that all the MACE etched surfaces reduce the reflectance quite significantly as compared to as-received wafers. Furthermore, the reflectance curves are relatively flat in all MACE samples having only a small peak around 600 nm. If we first consider only the micron-scale structures (i.e. ρ values 93-97%) a clear trend in the reflectance is observed so that the lower the $H_2O_2$ concentration, the lower the reflectance. Thus, the lowest average reflectance (15.3%) is obtained with a ρ value of 97%. The trend can be explained by comparing the results with the surface structure morphologies (Figure 1). There seems to be a correlation between the structure diameter and reflectance so that the larger the structure diameter, the higher the reflectance. If we then focus our attention to the sample that had clearly different surface morphology (nanometer-scale dimensions, Figure 1(a)), we can see that it has different optical properties as compared to the micron-scale counterparts. The sample has the lowest reflectance in the UV (presumably because the depth of the nanostructures is close to the dimensions of the photon wavelengths) while in the IR the reflectance is much higher and approaches 25% likely due to the shallow nature of the nanostructures.

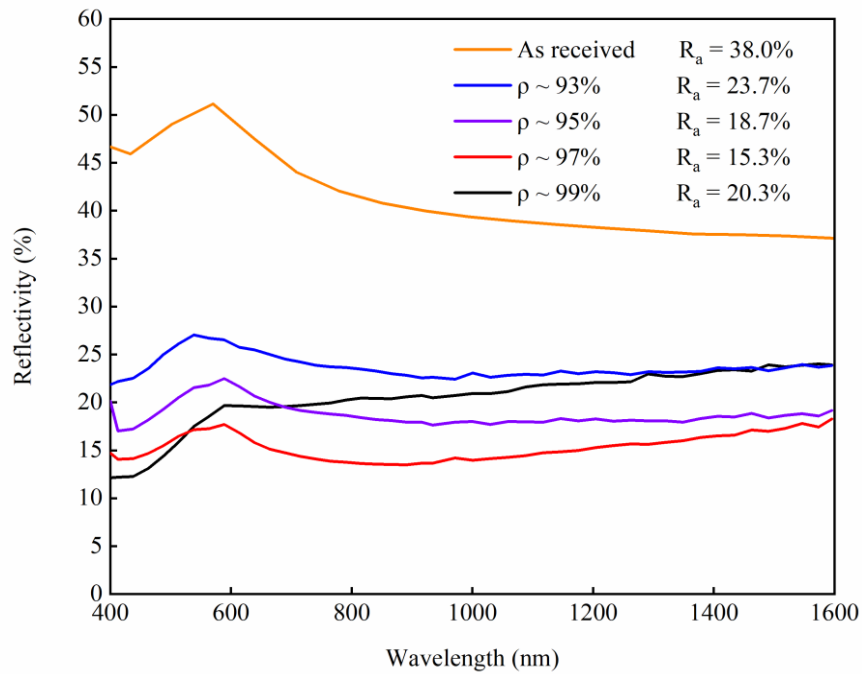

Figure 2. Reflectance spectra of nanostructured Ge samples etched for 3 min in 1-step MACE solution with different ρ values. The as-received wafer experiencing no MACE treatment is shown as a reference. Average reflectance over the presented wavelength range for every sample is reported in the legend.

In overall, all the reflectances reported above show only rather modest improvements as compared to state of the art [22] and are thus clearly not sufficient for most applications. The main reason for the modest reflectance is probably the shallow nature of the structures as already speculated above. More specifically, for instance Branz et. al. [32] have reported that reflection losses (especially at longer wavelengths) can be suppressed by forming deeper nanostructures. Such deep nanostructures are often obtained by proper tuning of the etching time [16].

In the above experiments, the etching time was only 3 minutes, therefore, it makes sense to study if the reflectance can be reduced further by prolonging the etching time. For the first prolonged etching time experiments we selected the sample with the ρ value of 97% since it provided clearly the best average reflectance after 3 min etching. Even though the diameter of the surface structure was already relatively long ~ 900 nm after 3min, the longer etching time could make the structure deeper, perhaps narrower, and thereby reduce the reflectance further. However, as shown in Figure 3, the reflectance is not lowered by increasing the etching time, on the contrary, the longest etching time (10 min) results in the worst reflectance. Thus, it seems that the mere increase in etching time is not the key in the process optimization and the deep and narrow nanostructures that should be ideally formed by MACE are still not achieved. In general, this result suggests that the ρ value of 97% is not the optimal concentration as it seems that $H_2O_2$ is still too aggressively attacking Ge surfaces. Therefore, lower concentration of $H_2O_2$ producing smaller initial structures was considered as a next logical attempt for making the structures deeper and narrower.

Based on the above, we next focused on the sample with nano-scale structure (Figure 1(a)), which was fabricated using the ρ value of 99%. As mentioned above, the depth of the structures could be likely enhanced by increasing the etching time but simultaneously one should avoid too much lateral extension of the structures due to uncontrolled oxidation and etching. Based on the SEM images shown in Figure 4(a), increasing the etching time in this sample indeed seems to make the structures much deeper (the depth increases from 140 nm to 230 nm) but without jeopardizing the nano-scale lateral dimensions that stay around 150 nm. Since the concentration of $H_2O_2$ is very low in this case, the generation of holes – and thus oxidation - takes most likely place only in the vicinity of the metal nanoparticles. Furthermore, since the HF concentration is high enough, it secures that the holes are confined and consumed near the nanoparticles, preventing their diffusion to larger areas. As the etching time proceeds, the reaction is limited close to the Ag nanoparticles, which then resembles the ideal MACE process mentioned in the introduction. However, when the etching time is increased further to 12 min, the structures start to show again lateral widening and the diameter of the structures extends all the way to 290 nm. This indicates that after reaching a specific point the direct reaction between HF/ $H_2O_2$ and Ge without Ag would start to dominate again over the conventional and ideal MACE process. This could be due to the earlier reported observation that after long enough etching time Ag is already largely consumed [33,34] and this causes the spreading of the structure dimensions despite the fact that Ag is known to be a relatively stable nanoparticle (e.g. as compared to Cu).

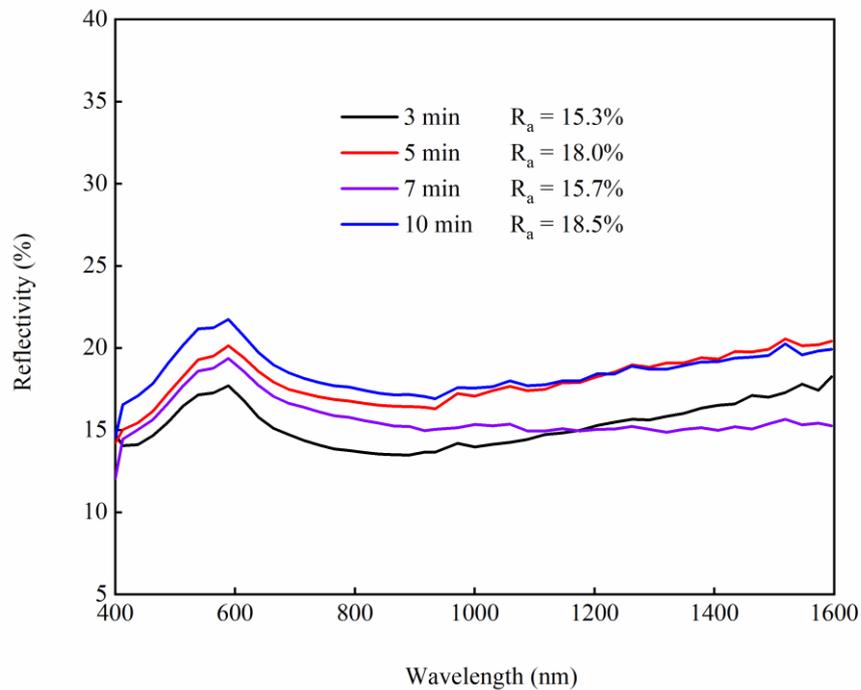

Figure 3. Reflectance spectra of nanostructured Ge samples after 1-step MACE process with ρ value of 97% after different etching times. Average reflectance over the presented wavelength range for every sample is reported in the legend.

Alternatively, after long etching time Ag nanoparticles are located so deep in the nanostructures that it takes more time from the chemicals to reach the bottom of the valley and a competitive reaction nearby starts to take place.

Figure 4(b) shows the corresponding reflectance spectra for the same Ge samples. Contrary to the sample with high $H_2O_2$, an excellent improvement in the optical performance is obtained by increasing the etching time. The samples show a clear trend with increasing etching times from 3 min up to 10 min. The best average reflectance of 9.0% at the wavelength range of 400-1600 nm is achieved after 10 min etching time and is better than has been reported before in Ge MACE process [22]. When the etching time is increased further, reflectivity starts to increase again. This behavior agrees with the changes seen in the morphology and as discussed above, the lateral etching takes over the ideal MACE process when the etching time is long enough (see Figure 4(a), lower right corner). It is worth to note that a relatively low reflectivity of ~12% is also achieved near the IR region, which indicates that MACE process can be beneficial for the applications focusing on this wavelength region as well.

Generally, the main benefit of the proposed MACE process, as compared to traditional antireflection coating, is that it is applicable simultaneously to a broad wavelength range (from UV to NIR). While there have been other ways to make nanostructures to Ge, e.g. plasma dry-etching processes [35,36] the benefit of MACE, in addition to industrial scalability, is that it can be applied also to thin layers of Ge. For instance in RIE, typical etching depth can be several microns [35] while here in MACE it is much less, only a few hundred nm.

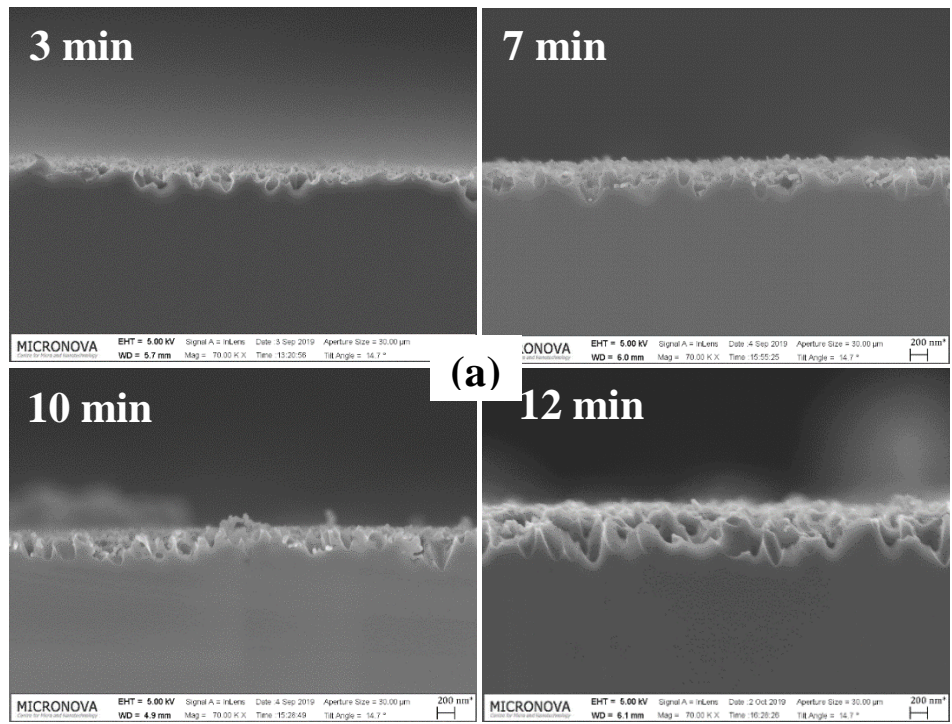

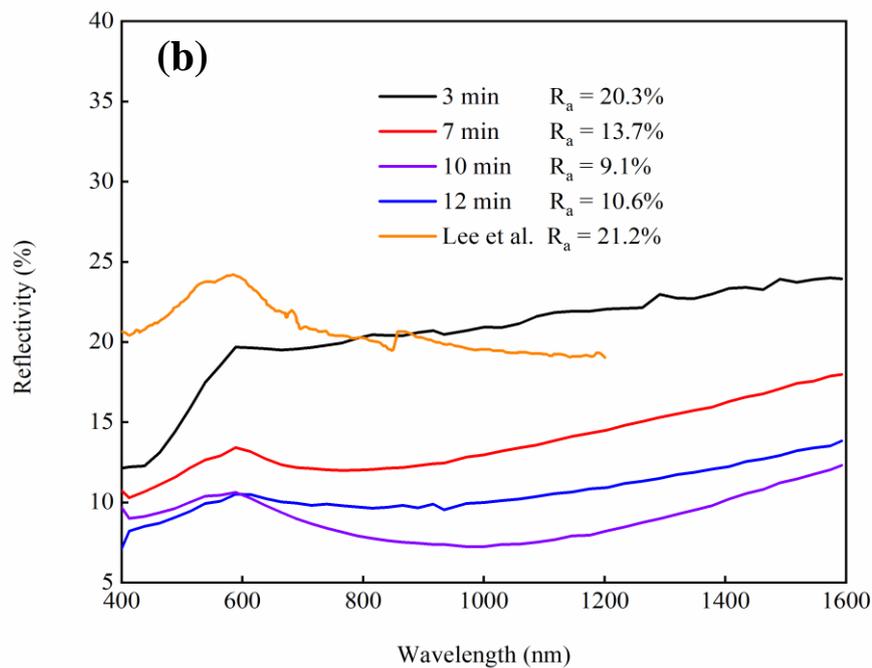

Figure 4. (a) SEM cross-sectional images and (b) the corresponding reflectance spectra of MACE nanostructured Ge samples with a ρ value of 99 % and with various etching times. The previous result of MACE Ge nanostructures by Lee et al. [22] is also shown as a reference. Average reflectance over the presented wavelength range for every sample is reported in the legend.

## 4. Conclusions

We have demonstrated that a simple one-step MACE process consisting of $AgNO_3$/HF/ $H_2O_2$ chemical mixture can be successfully applied to Ge surfaces, which results in a significant reduction in surface reflectance from over 40% to less than 10% over a wide wavelength range (400 - 1600 nm). We have shown that the concentration of $H_2O_2$ affects greatly the surface morphology and thus the optical properties of the surfaces. When the volume of $H_2O_2$ was increased, the obtained micron-scale structure resembled inverted cone-like shapes independent of the etching time. In contrast, a much lower concentration of $H_2O_2$ enabled nano-like pore formation that was further improved by extending the etching time. Consequently, the challenge related to the high reactivity of Ge surfaces could be overcome by properly tuning the etching parameters to confine the chemical reaction just under the Ag nanoparticles. Based on the results achieved here, it is safe to conclude that close to ideal MACE process is indeed possible also in Ge – and with performance similar than obtained earlier with Si.

## Acknowledgements

The authors acknowledge the provision of facilities by Aalto University at OtaNano – Micronova Nanofabrication Centre. Nicklas Anttu is acknowledged for the help in the optical characterization. The work was funded through the ATTRACT project funded by the EC under Grant Agreement 777222, by Business Finland through project RaPtor (687/31/2019) and by the Academy of Finland. The work is related to the Flagship on Photonics Research and Innovation "PREIN" funded by Academy of Finland.

## References

[1] Claeys C and Simoen E 2007 *Germanium-Based Technologies: From Materials to Devices* (Amsterdam: Elsevier Science)
[2] Hasegawa B H, Stebler B, Rutt B K, Martinez A, Gingold E L, Barker C S, Faulkner K G, Cann C E and Boyd D P 1991 A prototype high-purity germanium detector system with fast photon-counting circuitry for medical imaging *Med. Phys.* **18** 900–909
[3] Colace L and Assanto G 2009 Germanium on silicon for near-infrared light sensing *IEEE Photonics J.* **1** 69–79
[4] Soref R 2010 Mid-infrared photonics in silicon and germanium *Nat. Photonics* **4** 495–497
[5] Posthuma N E, Heide J V D, Flamand G and Poortmans J 2007 Emitter formation and contact realization by diffusion for germanium photovoltaic device *IEEE T. Electron Dev.* **54** 1210-1215
[6] Gryko J, McMillan P F, Marzke R F, Ramachandran G K, Patton D, Deb S K and Sankey O F 2000 Low-density framework form of crystalline silicon with a wide optical band gap *Phys. Rev. B* **62** R7707-R7710
[7] Park J H, Kuzum D, Jung W H and Saraswat K C 2011 Channel germanium MOSFET fabricated below 360°C by cobalt-induced dopant activation for monolithic three-dimensional-ICs *IEEE Electron Device Lett.* **32** 234-236
[8] Aalseth C E et al 2011 Results from a search for light-mass dark matter with a p-type point contact germanium detector *Phys. Rev. Lett.* **106** 131301
[9] Toriumi A and Nishimura T 2017 Germanium CMOS potential from material and process perspectives: Be more positive about germanium *Jpn. J. Appl. Phys.* **57** 010101
[10] Yeo C I, Kim J B, Song Y M and Lee Y T 2013 Antireflective silicon nanostructures with hydrophobicity by metal-assisted chemical etching for solar cell applications *Nanoscale Res. Lett.* **8** 159
[11] Li C, Zhao J, Yu X, Chen Q, Feng J and Sun H 2016 Fabrication of black silicon with thermostable infrared absorption by femtosecond laser *IEEE Photonics J.* **8** 6805809
[12] Gastrow G V, Alcubilla R, Ortega P, Yli-Koski M, Conesa-Boj S, Fontcuberta A, Morral I and Savin H 2015 Analysis of the atomic layer deposited $Al_2O_3$ field-effect passivation in black silicon *Sol. Energy Mater. Sol. Cells* **142** 29–33

[13] Chen K, Zha J, Hu F, Ye X, Zou S, Vähänissi V, Pearce J M, Savin H and Su X 2019 MACE nano-texture process applicable for both single- and multi-crystalline diamond-wire sawn Si solar cells *Sol. Energy Mater. Sol. Cells* **191** 1-8
[14] Chen K, Pasanen T P, Vähänissi V and Savin H 2019 Effect of MACE parameters on electrical and optical properties of ALD passivated black silicon *IEEE J. Photovolt.* **9** 974-979
[15] Toor F, Miller J B, Davidson L M, Nichols L, Duan W Q, Jura M P, Yim J, Forziat J and Black M R 2016 Nanostructured silicon via metal assisted catalyzed etch (MACE): chemistry fundamentals and pattern engineering *Nanotechnology* **27** 412003
[16] Huang Z P, Geyer N, Werner P, Boor J D and Gösele U 2011 Metal-assisted chemical etching of silicon: A review, *Adv. Mater.* **23** 285–308
[17] Ye X, Zou S, Chen K, Li J, Huang J, Cao F, Wang X, Zhang L, Wang X, Shen M and Su X 2014 18.45%-Efficient multi-crystalline silicon solar cells with novel nanoscale pseudo-pyramid texture *Adv. Funct. Mater.* **24** 6708–6716
[18] Cao F, Chen K, Zhang J, Ye X, Li J, Zou S and Su X 2015 Next-generation multi-crystalline silicon solar cells: diamond-wire sawing, nano-texture and high efficiency *Sol. Energy Mater. Sol. Cells* **141** 132–138
[19] Han H, Huang Z and Lee W 2014 Metal-assisted chemical etching of silicon and nano-technology applications, *Nano Today* **9** 271–304
[20] Zhang C, Li S Y, Ma W H, Ding Z, Wan X H, Yang J, Chen Z J, Zou Y X and Qiu J J 2017 Fabrication of ultra-low antireflection SiNWs arrays from mc-Si using one step MACE *J Mater Sci: Mater Electron* **28** 8510–8518
[21] Ito K, Yamaura D and Ogino T 2016 Chemical wet etching of germanium assisted with catalytic-metal-particles and electroless-metal-deposition *Electrochim. Acta* **214** 354-361
[22] Lee S, Choo H, Kim C, Oh E, Seo D and Lim S 2016 Metal-assisted chemical etching of Ge surface and its effect on photovoltaic devices *Appl. Surf. Sci.* **371** 129-138
[23] Kawase T, Mura A, Nishitani K, Kawai Y, Kawai K, Uchikoshi J, Morita M and Arima K 2012 Catalytic behavior of metallic particles in anisotropic etching of Ge (100) surfaces in water mediated by dissolved oxygen *J. Appl. Phys.* **111** 126102

[24] Kawase T, Mura A, Dei K, Nishitani K, Kawai K, Uchikoshi J, Morita M and Arima K 2013 Metal-assisted chemical etching of Ge (100) surfaces in water toward nanoscale patterning *Nanoscale Res. Lett.* **8** 151

[25] Torralba-Penalver E, Gall S L, Lachaume R, Magnin V, Harari J, Halbwax M, Vilcot J P, Cachet-Vivier C and Bastide S 2016 Tunable surface structuration of silicon by metal assisted chemical etching with Pt nanoparticles under electro-chemical bias, ACS *Appl. Mater. Interfaces* **8** 31375

[26] Yae S, Kawamoto Y, Tanakaa H, Fukumuro N and Matsuda H 2003 Formation of porous silicon by metal particle enhanced chemical etching in HF solution and its application for efficient solar cells *Electrochem. Commun.* **5** 632–636

[27] Dawood M K, Tripathy S, Dolmanan S B, Ng T H, Tan H and Lam J 2012 Influence of catalytic gold and silver metal nanoparticles on structural, optical, and vibrational properties of silicon nanowires synthesized by metal-assisted chemical etching *J. Appl. Phys.* **112** 073509

[28] Chartier C, Bastide s and Lévy-Clément C 2008 Metal-assisted chemical etching of silicon in HF-$H_2O_2$ *Electrochim. Acta* **53** 5509–5516

[29] Venkatesan R, Arivalagan M K, Venkatachalapathy V, Pearce J M and Mayandi J 2018 Effects of silver catalyst concentration in metal assisted chemical etching of silicon, *Mater. Lett.* **221** 206–210

[30] Lu Y T and Barron A R 2014 Anti-reflection layers fabricated by a one-step copper-assisted chemical etching with inverted pyramidal structures intermediate between texturing and nanopore-type black silicon *J. Mater. Chem.* A **2** 12043–12052

[31] Rezvani S J, Pinto N and Boarino L 2016 Rapid formation of single crystalline Ge nanowires by anodic metal assisted etching *CrystEngComm.* **18** 7843−7848

[32] Branz H M, Yost V E, Ward S, Jones K M, To B and Stradins P 2009 Nanostructured black silicon and the optical reflectance of graded-density surfaces *Appl. Phys. Lett.* **94** 231121

[33] Li S, Ma W, Zhou Y, Chen X, Xiao Y, Ma M, Zhu W and Wei F 2014 Fabrication of porous silicon nanowires by MACE method in $HF/H_2O_2/AgNO_3$ system at room temperature *Nanoscale Res. Lett.* **9** 196

[34] Lee C L, Tsujino K, Kanda Y, Ikeda S and Matsumura M 2008 Pore formation in silicon by wet etching using micrometre-sized metal particles as catalysts *J. Mater. Chem.* **18** 1015–1020

[35] Pasanen T P, Isometsä J, Garin M, Chen K, Vähänissi V and Savin H 2020 Nanostructured germanium with >99% absorption at 300–1600 nm wavelengths *Adv. Optical Mater.* **8** 2000047

[36] Steglich M, Käsebier T, Kley E B and Tünnermann A 2016 Black germanium fabricated by reactive ion etching *Appl. Phys. A* **122** 836